\documentclass{article}
\usepackage{dcase2019,amsmath,amssymb,graphicx,url,times,booktabs,tabularx,bm,multirow,bbm,todonotes}

\DeclareMathOperator*{\argmax}{arg\,max}

\title{SOUND EVENT LOCALIZATION AND DETECTION USING CRNN ON PAIRS OF MICROPHONES}

\twoauthors
  {Fran\c{c}ois Grondin, James Glass\sthanks{This work was funded in part by Signify.}}
    {Massachusetts Institute of Technology\\
CSAIL, 32 Vassar St.\\
     Cambridge, MA 02139, USA \\
     \{fgrondin,glass\}@mit.edu}
  {Iwona Sobieraj, Mark D. Plumbley\sthanks{The research leading to these results has received funding from the European Union's H2020 Framework Programme (H2020-MSCA-ITN-2014) under grant agreement n° 642685 MacSeNet. MDP is also supported by EPSRC grant EP/N014111/1.}}
    {University of Surrey \\
     Dept. Elec. \& Elec. Eng., 388 Stag Hill  \\
     Guildford, Surrey, GU2 7XH, UK \\
     \{i.sobieraj,m.plumbley\}@surrey.ac.uk}

\begin{document}

\ninept
\maketitle

\begin{sloppy}

\begin{abstract}
This paper proposes sound event localization and detection methods from multichannel recording. The proposed system is based on two Convolutional Recurrent Neural Networks (CRNNs) to perform sound event detection (SED) and time difference of arrival (TDOA) estimation on each pair of microphones in a microphone array.
In this paper, the system is evaluated with a four-microphone array, and thus combines the results from six pairs of microphones to provide a final classification and a 3-D direction of arrival (DOA) estimate.
Results demonstrate that the proposed approach outperforms the DCASE 2019 baseline system.
\end{abstract}

\begin{keywords}
Sound event detection, sound source localization, time difference of arrival, neural network
\end{keywords}

\section{Introduction}
\label{sec:intro}
Sound Event Detection (SED) is an important machine listening task, which aims to automatically recognize, label, and estimate the position in time of sound events in a continuous audio signal.
This is a popular research topic, due to the number of real-world applications for SED such as home-care \cite{hengel2009audio}, surveillance \cite{kotus2014detection}, environmental monitoring \cite{stowell2016} or urban traffic control \cite{meucci2008real}, to name just a few.
Successful Detection and Classification of Acoustic Scenes and Events (DCASE) challenges \cite{mesaros2018detection, mesaros2017dcase} now provide the community with datasets and baselines for a number of tasks related to SED.
However, most of the effort so far has concentrated on classification and detection of the sound events in time only, with little work done to perform robust localization of sound event in space.

Early approaches for SED are strongly inspired by speech recognition systems, using mel frequency cepstral coefficients (MFCCs) with Gaussian Mixture Models (GMMs) combined with Hidden Markov Models (HMM) \cite{heittola2013context,diment2013sound}.
Methods based on dictionary learning, mainly Non-negative Matrix Factorization (NMF), are also considered as prominent solutions for the SED task \cite{cotton2011spectral, komatsu2016acoustic, dikmen2013sound}.
With the recent advancements in machine learning, deep learning methods now provide state of the art results for this task \cite{parascandolo2016recurrent,xu2018large}.
The prevailing architectures used for SED are Convolutional Neural Networks (CNNs) \cite{Inoue2018}, which are particularly successful in computer vision tasks.
Other common approaches try to model time relations in audio signal by using recurrent neural networks (RNNs) \cite{parascandolo2016recurrent}.
Both can be combined in a Convolutional Recurrent Neural Network (CRNN), which achieves state of the art results on several machine listening tasks \cite{lu2018dcase2018, adavanne2018direction, adavanne2018sound}.

On the other hand, sound source localization (SSL) refers to estimating the direction of arrival (DOA) of multiple sound sources.
There are two popular categories of SSL methods: 1) high resolution and 2) steered-response techniques.
High resolution methods include Multiple Signal Classification (MUSIC) \cite{schmidt1986multiple} and Estimation of Signal Parameters via Rotational Invariance Technique (ESPRIT) \cite{roy1986estimation}.
These approaches, although initially designed for narrowband signals, can be adapted to broadband signals such as speech \cite{ishi2009evaluation, nakamura2012realtime, teutsch2005ebesprit, argentieri2007broadband, danes2010information}.
Alternatively, the Steered-Response Power Phase Transform (SRP-PHAT) robustly estimates the direction of arrival of speech and other broadband sources \cite{dibiase2001robust}.
SRP-PHAT relies on the Generalized Cross-Correlation with Phase Transform (GCC-PHAT) between each pair of microphones of a microphone array.
It is therefore convenient to estimate the time difference of arrival (TDOA) values for each pair, and combine these results to estimate the direction of arrival for a source \cite{grondin2013manyears, valin2007robust, valin2004localization, valin2006robust, grondin2019lightweight}, which is the approach we choose for this challenge.

In this paper we propose a system for sound event detection and localization (SELD), which we submitted to Task3 of the DCASE2019 Challenge \cite{adavanne2019dcase}.
Motivated by the results obtained by \cite{cao2019polyphonic}, we propose a CRNN architecture that uses both the spectrogram and GCC-PHAT features to perform SED and estimate TDOA.
However, since TDOA and SED have different cost functions, we believe they are distinct tasks with different optimal solutions, and we propose to use two separate neural networks for each of these two tasks. The results are then combined together to generate a final SED decision and estimate the DOA.

\section{Sound Event Localization and Detection}
\label{sec:task}
The goal of sound event localization and detection (SELD) is to output all instances of the sound events in the recording, its respective onset-offset times, and spatial locations in azimuth and elevation angles, given a multichannel audio input. An example of such a setup has been provided in Task 3 of the DCASE 2019 Challenge \cite{adavanne2019dcase}. Our system uses the TAU Spatial Sound Events - Microphone Array dataset, which provides four-channel directional microphone recordings from a tetrahedral array configuration. A detailed description of the dataset and the recording procedure may be found in \cite{adavanne2018sound}. 
In our approach, we propose to predict events and TDOAs for each pair of microphones, which leads to a total six pairs.

\section{Proposed method}
\label{sec:method}

We propose a method based on a combination of two convolutional recurrent neural networks (CRNNs), that share a similar front-end architecture. The first network, $\textrm{CRNN}_{SED}$, is trained to detect, label and estimate onset and offsets of sound events from a pair of microphones. The second network, $\textrm{CRNN}_{TDOA}$, estimates the TDOA for each pair of microphones and each class of sound events.
The SED results of all pairs are then combined together and a threshold is applied to make a final decision
regarding sound detection for each class.
The TDOAs are also combined together for all pairs of microphones and a DOA is generated for each class.
To obtain a DOA from the TDOA values, each potential DOA is assigned a set of target TDOAs, which are found during a initial calibration procedure.
Figure \ref{fig:overall} shows the overall architecture of the proposed system.
The following subsections describe in details each building block of the system.
\begin{figure}[!ht]
    \centering
    \includegraphics[width=\columnwidth]{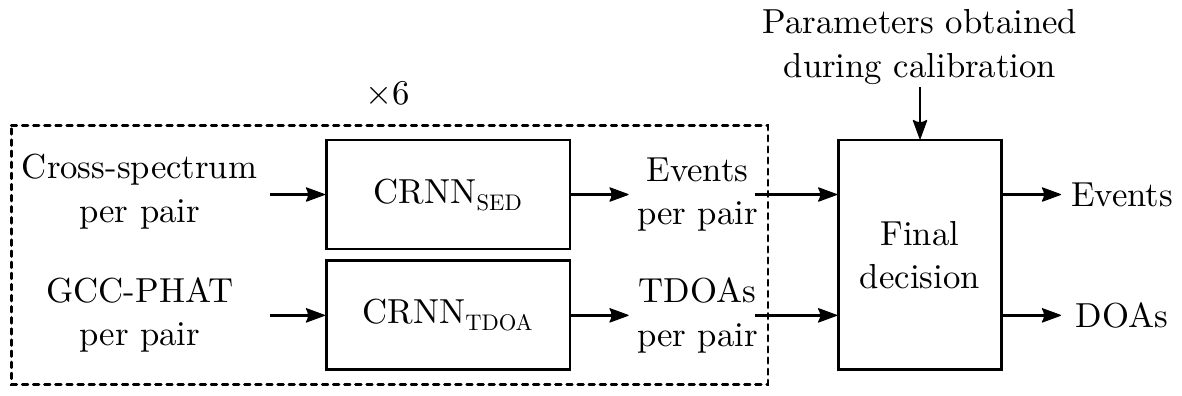}
    \vspace{-8pt}
    \caption{Architecture of the proposed system.}
    \label{fig:overall}
\end{figure}

\subsection{Calibration}

The search space around the microphone array is discretized into $Q$ DOAs, which are indexed by $q \in \mathcal{Q} = \{1, 2, \dots, Q\}$.
Each DOA $q$ is associated to an azimuth and an elevation, denoted by $(\phi_q, \theta_q)$, where $\phi_q \in \{-180^{\circ}, -170^{\circ}, \dots, +170^{\circ}\}$ and $\theta_q \in \{-40^{\circ}, -30^{\circ}, \dots, +40^{\circ}\}$, which corresponds to the discrete angles used when recording the DCASE dataset \cite{adavanne2019dcase}.  
The number of microphones corresponds to $M \in \mathbb{N}$, and the number of pairs to $P \in \mathbb{N}$, where $P = M(M-1)/2$.
Each DOA $q$ also corresponds to a vector $\bm{\tau}^q \in \mathcal{D}^P$ of TDOA values, where $\mathcal{D} = \{-\tau_{\max},\dots,+\tau_{\max}\}$ and the cardinality $|\mathcal{D}| = G$.
The expressions $\tau_{\max} \in \mathbb{R}^+$ and $G \in \mathbb{N}$ stand for the maximum TDOA and the number of discrete TDOA values, respectively.
Assuming free field propagation of sound, the microphone array geometry and the speed of sound provide enough information to estimate the TDOA values of each DOA.
However, the free field assumption becomes inaccurate when dealing with a closed microphone array (e.g. when microphones are installed around a filled support), and thus calibration based on the recorded signals is needed and is performed offline.

The expression $X^t_m[k] \in \mathbb{C}$ stands for the Short-Time Fourier Transform (STFT) coefficient at frame index $t \in \mathbb{N}$, microphone index $m \in \mathcal{M} = \{1,2,\dots,M\}$ and bin index $k \in \mathcal{K} = \{K_{\min},K_{\min}+1,\dots,K_{\max}\}$, where $K = K_{\max} - K_{\min}$ stands for the total number of frequency bins used.
The frame size and hop size correspond to $N \in \mathbb{N}$ and $\Delta N \in \mathbb{N}$, respectively, and the spectral content thus spans frequencies in the interval $[K_{\min}f_S/N,K_{\max}f_S/N]$ Hz, where $f_S \in \mathbb{R}_+$ stands for the sample rate in samples/sec.
The complex cross-spectrum $X^{t}_{i,j}[k]$ for each microphone pair $(i,j) \in \mathcal{P}=\{(x,y) \in \mathcal{M}^2: x<y\}$ corresponds to:
\begin{equation}
    \label{eq:cross-corr}
    X_{i,j}^{q,\mathcal{T}_q}[k] = \sum_{t \in \mathcal{T}_q}{X^t_i[k]X^t_j[k]^*}
\end{equation}
where $\mathcal{T}_q$ is a set that contains all the frame indexes where a single source is active at DOA $q$, and $(\dots)^*$ stands for the complex conjugate operator.
The Generalized Cross-Correlation with Phase Transform (GCC-PHAT) is then computed as follows:
\begin{equation}
    \label{eq:gcc-phat}
    x_{i,j}^{q,\mathcal{T}_q}[\tau] = \sum_{k \in \mathcal{K}}{W_{i,j}[\tau,k]\frac{X_{i,j}^{q,\mathcal{T}_q}[k]}{|X_{i,j}^{q,\mathcal{T}_q}[k]|}}
\end{equation}
where $W_{i,j}[\tau,k] = \exp(2\pi\sqrt{-1}\tau k/N)$, with $\tau \in \mathcal{D}$.

The TDOA value for the pair $(i,j)$ and DOA $q$ is then estimated as:
\begin{equation}
    \bar{\tau}^q_{i,j} = \argmax_{\tau \in \mathcal{D}}{\{x_{i,j}^{q,\mathcal{T}_q}[\tau]\}}.
\end{equation}
Since there is a limited amount of sound events per DOA in the training dataset, the estimated TDOAs $\bar{\tau}^q_{i,j}\,\forall\,(i,j)\,\in\,\mathcal{P}, \forall\,q\,\in\,\mathcal{Q}$ can be noisy.
To cope with this limitation, we apply a polynomial fitting method with an order of $27$ (found empirically).
For each discrete elevation angle $\theta \in \{-40^{\circ}, -30^{\circ}, \dots, +40^{\circ}\}$, there are $36$ azimuths $\phi \in \{-180^{\circ}, -170^{\circ}, \dots, +170^{\circ}\}$, and the TDOAs associated to these azimuths vary smoothly.
Therefore, for each pair $(i,j)$ and elevation $\theta$, we concatenate the estimated TDOAs three times to create a signal that spans over the azimuths $\phi \in \{-540^{\circ}, -530^{\circ}, \dots, +540^{\circ}\}$ and avoids the discontinuities observed at $-180^{\circ}$ and $170^{\circ}$ within the initial range.
A first polynomial fitting is then performed, and the outliers are removed prior to performing a second fitting, which finally provides the estimated TDOA $\tau^q_{i,j}$ for each DOA $q$ for the pair $(i,j)$:
\begin{equation}
    \tau^q_{i,j} = \textrm{polyfit}(\bar{\tau}^q_{i,j},27).
\end{equation}

Figure \ref{fig:calibration_polyfit} shows an example of the proposed method and how it deals effectively with outliers.
Note that once the polynomial coefficients are obtained, the TDOAs are only estimated in the region of interest, which is in the range $\phi \in \{-180^{\circ}, -170^{\circ}, \dots, +170^{\circ}\}$.
\begin{figure}[!ht]
    \centering
    \includegraphics[width=\columnwidth]{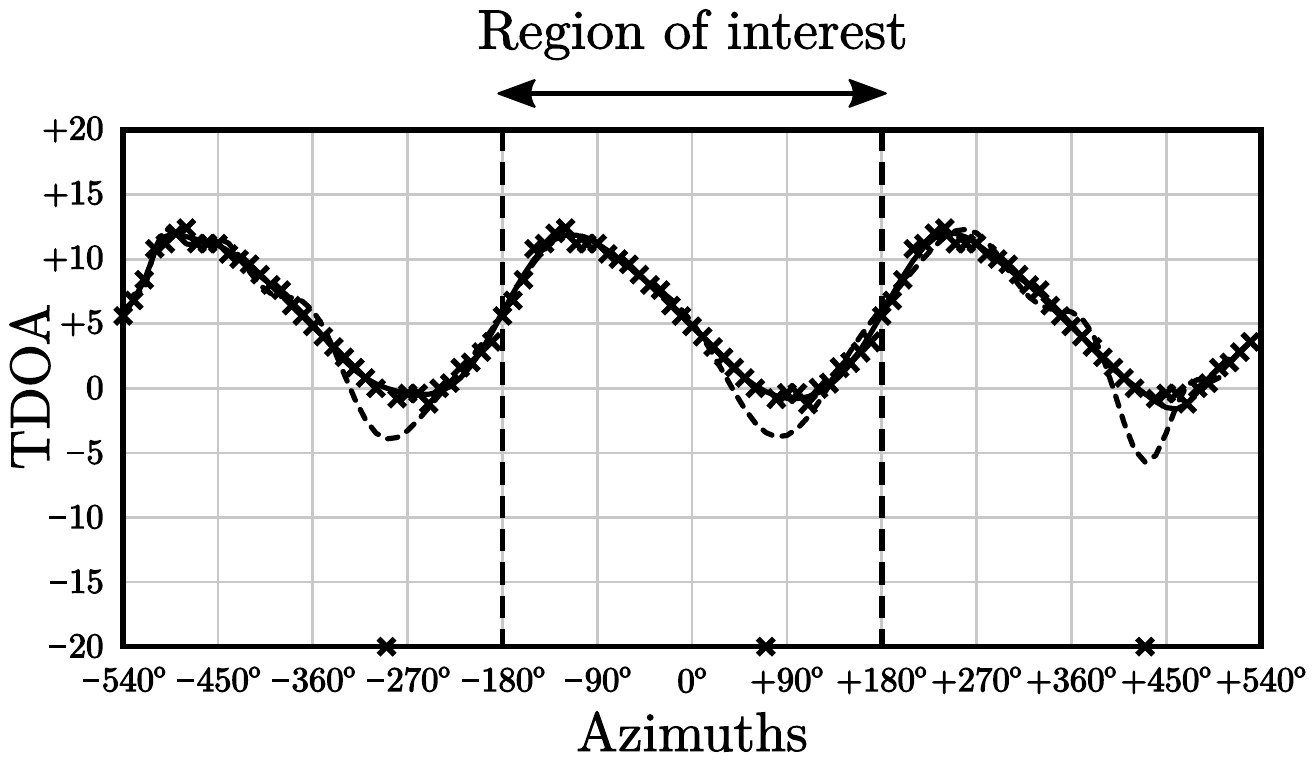}
    \caption{Calibration model for DOA estimation. First polynomial fit is shown as a dashed line, and the second one after removing the outliers is a solid line.}
    \label{fig:calibration_polyfit}
\end{figure}

\subsection{Neural network architecture}

The main building block of our system are two CRNNs that share a similar front-end architecture, as shown in Fig. \ref{fig:nn}. 

\begin{figure}[!ht]
    \centering
    \includegraphics[width=\columnwidth]{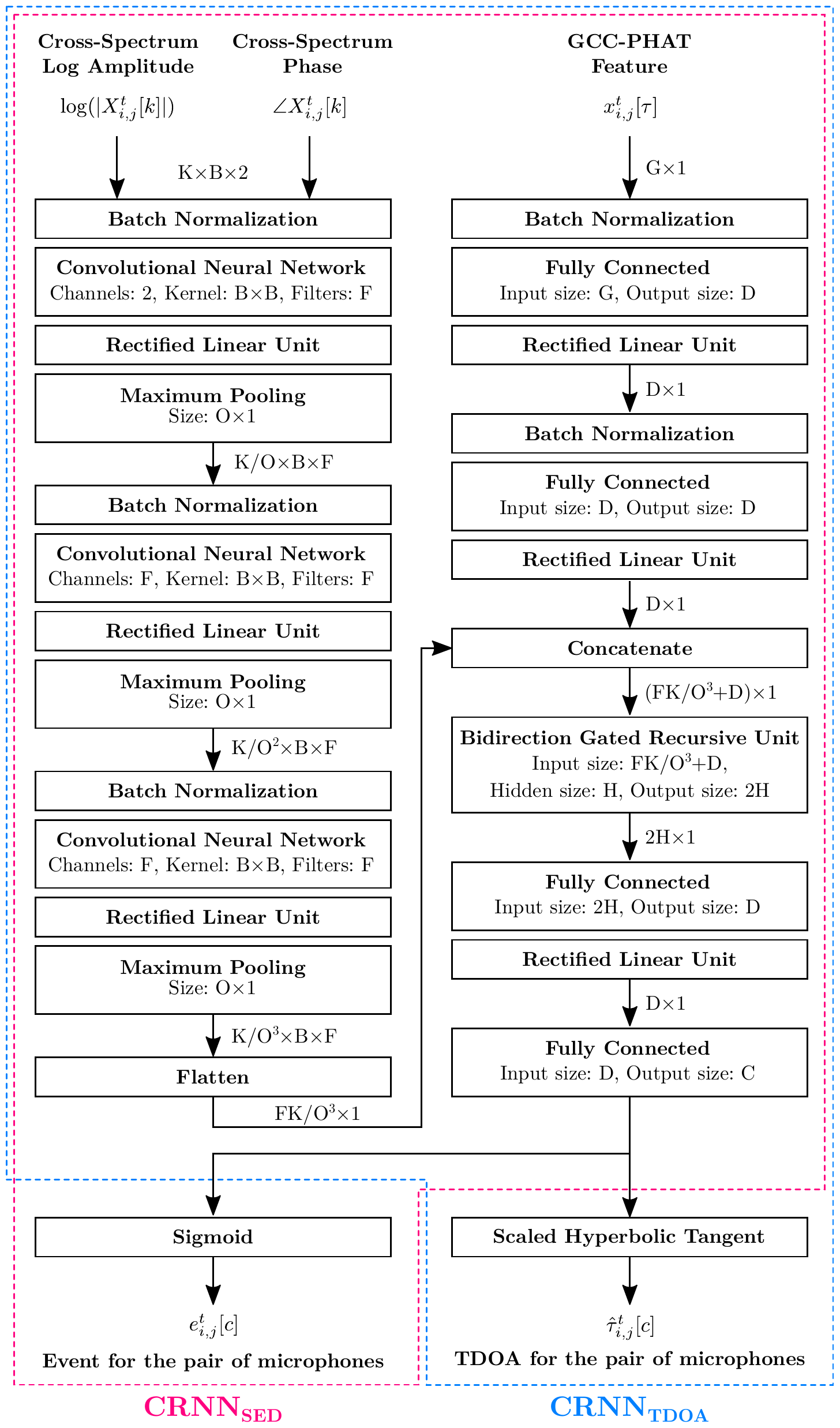}
    \caption{Architecture of $\textrm{CRNN}_{SED}$ and $\textrm{CRNN}_{TDOA}$}
    \label{fig:nn}
\end{figure}

The network consists of two branches.
This first is a series of convolutional layers (CNN), that process the log amplitude and phase of the instantaneous complex cross-spectrum input spectrograms (as in (\ref{eq:cross-corr})) between microphones $i$ and $j$. In parallel, GCC-PHAT features (as in (\ref{eq:gcc-phat}), but for a single frame $t$) are fed into a branch of a network that consists of two feed-forward layers. The outputs of two branches are concatenated and passed to a Bidirectional Gated Recurrent Unit (Bi-GRU) layer. The resultant vector is considered as a task dependent embedding of the input data. The embedding is passed to two feed forward layers, followed by an activation function, which depends on the task of the network.

$\textrm{CRNN}_{SED}$ is trained in a supervised manner using SED labels, i.e. information about the onset, offset and label of a sound event. As SED task may be pinned down to a multi-label classification of time frames, we use binary cross entropy as a loss function of the network. A Sigmoid activation function outputs the probabilities between $0$ and $1$ of each class for each time frame.

$\textrm{CRNN}_{TDOA}$ is trained on TDOA labels for each pair of microphones. The problem of TDOA estimation is defined in a regression framework.
Hence, Mean Squared Error (MSE) loss is used to train the network.
Similarly to the $\textrm{CRNN}_{SED}$, the network consists of CNNs and GRU, followed by an activation function, Hyperbolic Tangent (tanh) in this case, scaled by $\tau_{\max}$ as the TDOA value lies in the range $[-\tau_{\max},+\tau_{\max}]$. 
Note that the TDOA is only estimated over segments (i.e. audio samples for a given time interval) where the corresponding sound event is active according to the reference labels, as proposed in \cite{cao2019polyphonic}.

Both networks are trained separately on all pairs of microphones, using segments of $3$ seconds selected randomly amongst the training dataset, and using the Adam optimizer with a learning rate of $10^{-3}$ and a batch size of $16$.
We stopped training the network when no further improvement is observed on the validation set, that is after 120,000 segments for $\textrm{CRNN}_{SED}$ and 160,000 segments for $\textrm{CRNN}_{TDOA}$.

\subsection{Event detection}

$\textrm{CRNN}_{SED}$ returns a value $e^t_{i,j}[c] \in [0,1]$ for each pair of microphones $(i,j)$ and class $c \in \{1, 2, \dots, C\}$.
These values are summed up for all pairs and each class, and normalized by the number of pairs, which leads to a new expression $e^t[c] \in [0,1]$:
\begin{equation}
    e^t[c] = \frac{1}{P}\sum_{i=1}^{M}\sum_{j=i+1}^{M}{e_{ij}^t[c]}.
\end{equation}
An event from class $c$ is then considered to be detected at frame $t$ if $e^t[c]$ exceeds a threshold, which is class specific:
\begin{equation}
    E^t[c] = \begin{cases}
    1 & e^t[c] \geq \epsilon[c] \\
    0 & e^t[c] < \epsilon[c] \\
    \end{cases}.
\end{equation}
A post-filter method finally ensures that each sound event lasts a minimum amount of frames (denoted by $\gamma$) to avoid  
false detection of sporadic events.
For evaluation purpose, the event activity is usually defined for a given segment $l$, where $\mathcal{T}^l = \{tL, tL+1, \dots, t(L+1)-1\}$ holds the $L$ frames that belong to segment $l$.
The estimated event activity $Event_E^l[c]$ is then said to be active if at least one frames within the interval indicates the event is active.

\subsection{DOA estimation}

Similarly to $\textrm{CRNN}_{SED}$, $\textrm{CRNN}_{TDOA}$ returns an estimated TDOA $\hat{\tau}^t_{i,j}[c]$ for each class $c$ and pair of microphone $(i,j)$ at frame $t$.
For each DOA at index $q$, the estimated TDOAs $\hat{\tau}^t_{i,j}[c]$ are compared to the theoretical values $\tau^q_{i,j}$ obtained from polynomial fitting during the calibration step.
A Gaussian kernel with a variance of $\sigma^2$ then generates a value close to $1$ when both TDOAs are close to each other, whereas this value goes to zero when the difference increases.
All DOAs are scanned for each class, and the one that returns the maximum sum corresponds to the estimated DOA index $q^t[c]^*$: 
\begin{equation}
    q^t[c]=\argmax_{q\,\in\mathcal{Q}}\sum_{i=1}^{M}\sum_{j=i+1}^{M}{\exp\left[\frac{\left(\hat{\tau}_{ij}^t[c]-\tau_{ij}^q\right)^2}{2\sigma^2}\right]}.
\end{equation}
The estimated DOAs are then concatenated in $\mathbf{DOA}^t_E$:
\begin{equation}
    \mathbf{DOA}_E^t = \{ (\phi_{q^t[c]},\theta_{q^t[c]}) \}\ \forall\ c\ \textrm{where}\ \hat{E}^t[c] = 1.
\end{equation}

\section{Results}
\label{sec:results}

The proposed system is evaluated on the DCASE 2019 development dataset.
This set is divided into 4 cross-validation splits of 100 one-minute recordings each, as described in \cite{adavanne2018sound}.
Table \ref{tab:results_parameters} lists the parameters used in the experiments.
The sample rate $f_S$ and the number of microphones $M$ match the DCASE dataset parameters.
The frame size $N$ corresponds to 43 msecs, which allows a good trade-off between time and frequency resolutions.
The hop size $\Delta N$ provides a spacing of $20$ msecs between frames, which corresponds to the hop length for evaluation in the actual challenge.
The values of $K_{\min}$ and $K_{\max}$ are set to provide a frequency range that goes up to 12 kHz (and exclude the DC component), which is where most of the sound event energy lies.
The parameter $\gamma$ is chosen to ensure a minimum sound event duration of $100$ msecs, and the standard deviation $\sigma$ is found empirically to provide a good DOA resolution with $G$ TDOA values.
The maximum value for a TDOA is set such that this includes all possible TDOA values for the actual array geometry.
Finally, the neural network hyperparameters $B$, $F$, $O$, $H$ and $D$ are found empirically from observed performances with the validation set.
Also note that the event thresholds $\epsilon[c]$ are found empirically by scanning values between $0$ and $1$ and selecting thresholds that lead to the best event detection metrics on the validation set.

\begin{table}[ht]
    \centering
    \renewcommand{\arraystretch}{1.05}
    \begin{tabular}{|cc|c|cc|c|cc|}
        \cline{1-2}\cline{4-5}\cline{7-8}
        Param. & Value & & Param. & Value & & Param. & Value \\
        \cline{1-2}\cline{4-5}\cline{7-8}
        $f_S$ & $48000$ & & $K_{\min}$ & $1$ & & $B$ & $3$ \\
        $M$ & $4$ & & $K_{\max}$ & $513$ & & $F$ & $64$ \\
        $N$ & $2048$ & & $\gamma$ & $5$ & & $O$ & $4$ \\
        $\Delta N$ & $960$ & & $\sigma$ & $2.0$ & & $H$ & $512$ \\
        $\tau_{\max}$ & $20.0$ & & $G$ & $101$ & & $D$ & $256 $\\
        \cline{1-2}\cline{4-5}\cline{7-8}
    \end{tabular}
    \caption{Parameters of the proposed system}
    \label{tab:results_parameters}
\end{table}

To evaluate the performance of the system, events are defined for segments of 1 sec ($L=50$).
We define the number of true positives ($TP^l$) for segment $l$ as the number of correctly estimated events with respect to the reference events activity ($Event^l_R[c]$):
\begin{equation}
    TP^l = \sum_{c=1}^{C}{Event^l_E[c] \cdot Event^l_R[c]}.
\end{equation}
Similarly, the number of false negatives ($FN^l$) and false positives ($FP^l$) are given by:
\begin{equation}
    FN^l = \sum_{c=1}^{C}{Event^l_E[c] \cdot(1 - Event^l_R[c])}
\end{equation}
\begin{equation}
    FP^l = \sum_{c=1}^{C}{(1-Event^l_E[c]) \cdot Event^l_R[c]}.
\end{equation}
Finally the total number of active events corresponds to:
\begin{equation}
    N^l = \sum_{c=1}^{C}{Event^l_R[c]}.
\end{equation}
We then define substitutions ($S^l$), deletions ($D^l$) and insertions ($I^l$) are defined as:
\begin{equation}
    S^l = \min{\{FN^l, FP^l\}}
\end{equation}
\begin{equation}
    D^l = \max{\{0, FN^l - FP^l\}}
\end{equation}
\begin{equation}
    I^l = \max{\{0, FP^l - FN^l\}}.
\end{equation}
This leads to the event rate (ER) and F1-score (F) metrics \cite{mesaros2016metrics}:
\begin{equation}
    ER = \frac{\sum_{l}{S^l} + \sum_{l}{D^l} + \sum_{l}{I^l}}{\sum_{l}{N^l}}
\end{equation}
\begin{equation}
    F = \frac{2\sum_l{TP^l}}{2\sum_l{TP^l} + \sum_l{FN^l} + \sum_l{FP^l}}.
\end{equation}
The DOA metrics consist of the DOA error (DOAE) and frame recall (FR) \cite{adavanne2018direction}.
The DOAE is obtained as follows:
\begin{equation}
    DOAE = \left(\sum_{t=1}^{T}{D^t_E}\right)^{-1}\sum_{t=1}^{T}{\mathcal{H}(\mathbf{DOA}_R^t,\mathbf{DOA}_E^t)}
\end{equation}
where $D_E^t$ denotes the number of estimated events, $\mathcal{H}(\dots)$ stands for Hungarian algorithm \cite{adavanne2018direction} and $\mathbf{DOA}^t_R$ represents the reference DOA.
The pair-wise costs between individual predicted and reference DOAs corresponds to:
\begin{equation}
    h = \arccos{(\sin{\phi_E}\sin{\phi_R} + \cos{\phi_E}\cos{\phi_R}\cos{(\theta_R-\theta_E)})}
\end{equation}
where $\phi_E$ and $\phi_R$ stand for the azimuth of the estimated and reference DOA, respectively, and $\theta_E$ and $\theta_R$ stand for the elevation of the estimated and reference DOA, respectively.

Finally, the frame recall corresponds to the following expression, where $D_R^t$ denotes the number of reference events, and $\mathbbm{1}(\dots)$ stands for the indicator function that generates an output one if the condition ($D^t_R = D^t_E$) is met, or zero otherwise:
\begin{equation}
    FR = \frac{1}{T}\sum_{t=1}^{T}{\mathbbm{1}(D^t_R = D^t_E)}.
\end{equation}
Table \ref{tab:results} summarizes the results for the baseline and the proposed method.
This shows that the proposed system outperforms the baseline for all metrics, and improves particularly the accuracy of the estimated DOA. 

\begin{table}[ht]
    \centering
    \renewcommand{\arraystretch}{1.05}
    \begin{tabular}{|cccccc|}
        \hline
        Method & Dataset & ER & F & DOAE & FR \\
        \hline
        \multirow{2}{*}{Baseline} & Dev. & $0.35$ & $80.0\%$ & $30.8^{\circ}$ & $84.0\%$ \\
        & Eval. & $0.28$ & $85.4\%$ & $24.6^{\circ}$ & $85.7\%$ \\
        \hline
        \multirow{2}{*}{Proposed} & Dev. & $\bm{0.21}$ & $\bm{87.2\%}$ & $\bm{6.8^{\circ}}$ & $\bm{84.7\%}$ \\
        & Eval. & $\bm{0.14}$ & $\bm{92.2\%}$ & $\bm{7.4^{\circ}}$ & $\bm{87.5\%}$ \\
        \hline
    \end{tabular}
    \caption{Performances in terms of Error Rate (ER -- less is better), F score (F -- more is better), Direction of Arrival Error (DOA -- less is better) and Frame Recall (FR -- more is better).}
    \label{tab:results}
\end{table}

\section{Conclusion}
\label{sec:conclusion}

In this paper, we propose a system to detect sound events and estimate their TDOA for each pair of microphones, which then combines them to detect sound events and estimate their DOA for a four-microphone array.
The proposed method outperforms the DCASE 2019 baseline system.

In future work, additional neural networks architecture should be investigated for SELD.
Moreover, making the system work online (by using unidirectional GRU layers for instance) would make the method appealing for real-world applications.

\bibliographystyle{IEEEtran}
\bibliography{refs}

\begin{thebibliography}{10}
\providecommand{\url}[1]{#1}
\def\UrlFont{\rmfamily}
\providecommand{\newblock}{\relax}
\providecommand{\bibinfo}[2]{#2}
\providecommand\BIBentrySTDinterwordspacing{\spaceskip=0pt\relax}
\providecommand\BIBentryALTinterwordstretchfactor{4}
\providecommand\BIBentryALTinterwordspacing{\spaceskip=\fontdimen2\font plus
\BIBentryALTinterwordstretchfactor\fontdimen3\font minus
  \fontdimen4\font\relax}
\providecommand\BIBforeignlanguage[2]{{%
\expandafter\ifx\csname l@#1\endcsname\relax
\typeout{** WARNING: IEEEtran.bst: No hyphenation pattern has been}%
\typeout{** loaded for the language `#1'. Using the pattern for}%
\typeout{** the default language instead.}%
\else
\language=\csname l@#1\endcsname
\fi
#2}}

\bibitem{hengel2009audio}
P.~van Hengel and J.~Anem{\"u}ller, ``Audio event detection for in-home care,''
  in \emph{Proc. ICA}, 2009, pp. 618--620.

\bibitem{kotus2014detection}
J.~Kotus, K.~Lopatka, and A.~Czyzewski, ``Detection and localization of
  selected acoustic events in acoustic field for smart surveillance
  applications,'' \emph{Multimed. Tools Appl.}, vol.~68, no.~1, pp. 5--21,
  2014.

\bibitem{stowell2016}
D.~Stowell, M.~Wood, Y.~Stylianou, and H.~Glotin, ``Bird detection in audio:
  {A} survey and a challenge,'' in \emph{Proc. IEEE MLSP}, 2016.

\bibitem{meucci2008real}
F.~Meucci, L.~Pierucci, E.~Del~Re, L.~Lastrucci, and P.~Desii, ``A real-time
  siren detector to improve safety of guide in traffic environment,'' in
  \emph{Proc. EUSIPCO}, 2008.

\bibitem{mesaros2018detection}
A.~Mesaros, T.~Heittola, E.~Benetos, P.~Foster, M.~Lagrange, T.~Virtanen, and
  M.~D. Plumbley, ``Detection and classification of acoustic scenes and events:
  Outcome of the {DCASE} 2016 challenge,'' \emph{IEEE/ACM Trans. Audio, Speech,
  Language Process.}, vol.~26, no.~2, pp. 379--393, 2018.

\bibitem{mesaros2017dcase}
A.~Mesaros, T.~Heittola, A.~Diment, B.~Elizalde, A.~Shah, E.~Vincent, B.~Raj,
  and T.~Virtanen, ``{DCASE} 2017 challenge setup: Tasks, datasets and baseline
  system,'' in \emph{Proc. DCASE Workshop}, 2017.

\bibitem{heittola2013context}
T.~Heittola, A.~Mesaros, A.~Eronen, and T.~Virtanen, ``Context-dependent sound
  event detection,'' \emph{EURASIP J. Audio, Spee.}, vol. 2013, no.~1, pp.
  1--13, 2013.

\bibitem{diment2013sound}
A.~Diment, T.~Heittola, and T.~Virtanen, ``Sound event detection for office
  live and office synthetic {AASP} challenge,'' in \emph{Proc. IEEE AASP
  DCASE}, 2013, pp. 1--3.

\bibitem{cotton2011spectral}
C.~V. Cotton and D.~P. Ellis, ``Spectral vs. spectro-temporal features for
  acoustic event detection,'' in \emph{Proc. IEEE WASPAA}, 2011, pp. 69--72.

\bibitem{komatsu2016acoustic}
T.~Komatsu, T.~Toizumi, R.~Kondo, and Y.~Senda, ``Acoustic event detection
  method using semi-supervised non-negative matrix factorization with a mixture
  of local dictionaries,'' in \emph{Proc. DCASE Workshop}, 2016, pp. 45--49.

\bibitem{dikmen2013sound}
O.~Dikmen and A.~Mesaros, ``Sound event detection using non-negative
  dictionaries learned from annotated overlapping events,'' in \emph{Proc. IEEE
  WASPAA}, 2013.

\bibitem{parascandolo2016recurrent}
G.~Parascandolo, H.~Huttunen, and T.~Virtanen, ``Recurrent neural networks for
  polyphonic sound event detection in real life recordings,'' in \emph{Proc.
  IEEE ICASSP}, 2016, pp. 6440--6444.

\bibitem{xu2018large}
Y.~Xu, Q.~Kong, W.~Wang, and M.~D. Plumbley, ``Large-scale weakly supervised
  audio classification using gated convolutional neural network,'' in
  \emph{Proc. IEEE ICASSP}, 2018, pp. 121--125.

\bibitem{Inoue2018}
T.~Inoue, P.~Vinayavekhin, S.~Wang, D.~Wood, N.~Greco, and R.~Tachibana,
  ``Domestic activities classification based on {CNN} using shuffling and
  mixing data augmentation,'' DCASE2018 Challenge, Tech. Rep., 2018.

\bibitem{lu2018dcase2018}
L.~JiaKai, ``Mean teacher convolution system for {DCASE} 2018 task 4,''
  DCASE2018 Challenge, Tech. Rep., 2018.

\bibitem{adavanne2018direction}
S.~Adavanne, A.~Politis, and T.~Virtanen, ``Direction of arrival estimation for
  multiple sound sources using convolutional recurrent neural network,'' in
  \emph{Proc. IEEE EUSIPCO}, 2018, pp. 1462--1466.

\bibitem{adavanne2018sound}
S.~Adavanne, A.~Politis, J.~Nikunen, and T.~Virtanen, ``Sound event
  localization and detection of overlapping sources using convolutional
  recurrent neural networks,'' \emph{J. Sel. Topics Signal Process.}, vol.~13,
  pp. 34--48, 2018.

\bibitem{schmidt1986multiple}
R.~Schmidt, ``Multiple emitter location and signal parameter estimation,''
  \emph{IEEE Trans. Antennas Propag.}, vol.~34, no.~3, pp. 276--280, 1986.

\bibitem{roy1986estimation}
R.~Roy, A.~Paulraj, and T.~Kailath, ``Estimation of signal parameters via
  rotational invariance techniques - {ESPRIT},'' in \emph{Proc. IEEE MILCOM},
  1986.

\bibitem{ishi2009evaluation}
C.~Ishi, O.~Chatot, H.~Ishiguro, and N.~Hagita, ``{Evaluation of a MUSIC-based
  real-time sound localization of multiple sound sources in real noisy
  environments},'' in \emph{Proc. IEEE/RSJ IROS}, 2009, pp. 2027--2032.

\bibitem{nakamura2012realtime}
K.~Nakamura, K.~Nakadai, and H.~Okuno, ``A real-time super resolution robot
  audition system that improves the robustness of simultaneous speech
  recognition,'' \emph{Adv. Robotics}, vol.~27, no.~12, pp. 933--945, 2013.

\bibitem{teutsch2005ebesprit}
H.~Teutsch and W.~Kellermann, ``{EB-ESPRIT}: {2D} localization of mulitple
  wideband acoustic sources using eigen-beams,'' in \emph{Proc. IEEE ICASSP},
  2005, pp. 89--92.

\bibitem{argentieri2007broadband}
S.~Argentieri and P.~Dan\`es, ``Broadband variations of the {MUSIC}
  high-resolution method for sound source localization in robotics,'' in
  \emph{Proc. IEEE/RSJ IROS}, 2007, pp. 2009--2014.

\bibitem{danes2010information}
P.~Dan\`es and J.~Bonnal, ``Information-theoretic detection of broadband
  sources in a coherent beamspace {MUSIC} scheme,'' in \emph{Proc. IEEE/RSJ
  IROS}, 2010, pp. 1976--1981.

\bibitem{dibiase2001robust}
J.~DiBiase, H.~Silverman, and M.~Brandstein, ``Robust localization in
  reverberant rooms,'' in \emph{Microphone Arrays}.\hskip 1em plus 0.5em minus
  0.4em\relax Springer, 2001, pp. 157--180.

\bibitem{grondin2013manyears}
F.~Grondin, D.~L{\'e}tourneau, F.~Ferland, V.~Rousseau, and F.~Michaud, ``The
  manyears open framework,'' \emph{Autonomous Robots}, vol.~34, no.~3, pp.
  217--232, 2013.

\bibitem{valin2007robust}
J.-M. Valin, F.~Michaud, and J.~Rouat, ``{Robust localization and tracking of
  simultaneous moving sound sources using beamforming and particle
  filtering},'' \emph{Rob. Auton. Syst.}, vol.~55, no.~3, pp. 216--228, 2007.

\bibitem{valin2004localization}
J.-M. Valin, F.~Michaud, B.~Hadjou, and J.~Rouat, ``Localization of
  simultaneous moving sound source for mobile robot using a frequency-domain
  steered beamformer approach,'' in \emph{Proc. IEEE ICRA}, 2004, pp.
  1033--1038.

\bibitem{valin2006robust}
J.-M. Valin, F.~Michaud, and J.~Rouat, ``Robust {3D} localization nad tracking
  of sound sources using beamforming and particle filtering,'' in \emph{Proc.
  IEEE ICASSP}, 2006, pp. 841--844.

\bibitem{grondin2019lightweight}
F.~Grondin and F.~Michaud, ``Lightweight and optimized sound source
  localization and tracking methods for open and closed microphone array
  configurations,'' \emph{Rob. Auton. Syst.}, vol. 113, pp. 63--80, 2019.

\bibitem{adavanne2019dcase}
S.~Adavanne, A.~Politis, and T.~Virtanen, ``A multi-room reverberant dataset
  for sound event localization and detection,'' in \emph{Submitted to DCASE
  Workshop}, 2019.

\bibitem{cao2019polyphonic}
Y.~Cao, Q.~Kong, T.~Iqbal, F.~An, W.~Wang, and M.~D. Plumbley, ``Polyphonic
  sound event detection and localization using a two-stage strategy,''
  \emph{arXiv preprint arXiv:1905.00268}, 2019.

\bibitem{mesaros2016metrics}
A.~Mesaros, T.~Heittola, and T.~Virtanen, ``Metrics for polyphonic sound event
  detection,'' \emph{Applied Sciences}, vol.~6, no.~6, p. 162, 2016.

\end{thebibliography}

\end{sloppy}
\end{document}